\documentclass[aps,prb,twocolumn,reprint,showpacs]{revtex4-1}
\usepackage{amsmath}
\usepackage{amssymb}
\usepackage{graphicx}
\usepackage{hyperref}
\usepackage{color,xcolor}
\usepackage{epstopdf}
\begin{document}
\title{Strain enhanced superconductivity of Mo$X_2$ ($X$=S or Se) bilayers with Na intercalation}
\author{Jun-Jie Zhang}
\author{Bin Gao}
\author{Shuai Dong}
\email{Corresponding author: sdong@seu.edu.cn}
\affiliation{Department of Physics, Southeast University, Nanjing 211189, China}
\date{\today}

\begin{abstract}
Mo$X_2$ ($X$=S or Se) is a semiconductor family with two-dimensional structure. And a recent calculation predicted the superconductivity in electron doped MoS$_2$ monolayer. In this work, the electronic structure and lattice dynamics of Mo$X_2$ bilayers with monolayer Na intercalated, have been calculated. According to the electron-phonon interaction, it is predicted that these bilayers can be transformed from indirect-gap semiconductors to a superconductors by Na intercalation. More interestingly, the biaxial tensile strain can significantly enhance the superconducting temperature up to $\sim10$ K in Na-intercalated MoS$_2$. In addition, the phonon mean free path at room-temperature is also greatly improved in Na intercalated MoSe$_2$, which is advantaged for related applications.
\end{abstract}
\maketitle

\section{Introduction}
Two-dimensional (2D) material such as graphene, black phosphorus, and layered transition metal dichalcogenides (TMDs) have attracted enormous interest for their unique structure, novel physical properties, and broad potential applications. Superior to the semi-metallic graphene, few-layer MoS$_2$'s show moderate band gaps which are crucial for practical on/off ratio in electronic circuit devices.\cite{radisavljevic2011single} Besides, MoS$_2$ is potentially important in optoelectronic because its band gap is in the visible light range. \cite{wang2012electronics} For this reason, great efforts have been made to investigate the dynamics of various carriers in MoS$_2$ including mobilities of electrons, excitons, as well as phonons. \cite{sim2013exciton,shi2013exciton,kaasbjerg2012phonon,kaasbjerg2013acoustic}

\begin{figure}
\centering
\includegraphics[width=0.48\textwidth]{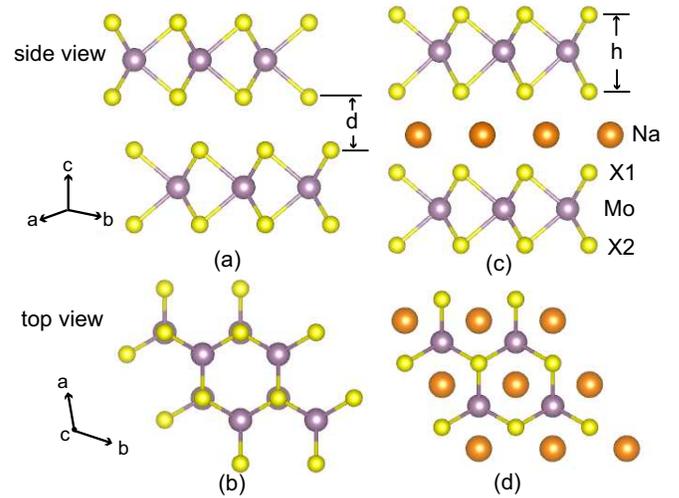}
\caption{(Color online) Side and top views of atomic structures of Mo$X_2$ bilayer [(a) and (b)] and (Mo$X_2$)$_2$Na system [(c) and (d)].}
\label{Fig1}
\end{figure}

Structurally, each layer of MoS$_2$ (as well as MoSe$_2$) is constructed by the S-Mo-S (or Se-Mo-Se) sandwich, as shown in Fig.~\ref{Fig1}. Along the $c$-axis, the neighboring triatomic layer is weakly coupled by van der Waals (vdw) interaction.\cite{verble1970lattice} This layered nature makes MoS$_2$ (or MoSe$_2$) flexible and tailorable, e.g. to be doped by ion absorption or intercalation, as well as to fabricate heterostructures. \cite{lee2015giant,huang2013density,ge2013phonon} These artificial modificatory MoS$_2$ and MoSe$_2$ show lots of extraordinary qualities. For example, a giant Rashba-type splitting was found in MoS$_2$/Bi(111) heterostructure, \cite{lee2015giant} and a half-metal behavior was predicted in Fe adatoms adsorbed on monolayer and bilayer MoS$_2$ sheets.\cite{huang2013density} Recently, the electron-doped monolayer MoS$_2$ was predicted to be a BCS-type superconductor with a considerable critical temperature ($T_{\rm C}$) up to $20$ K when one extra electron artificially-added per chemical unit.\cite{ge2013phonon} However, by considering realistic doping methods (e.g. K absorption),\cite{ge2013phonon} the calculated $T_{\rm C}$ was far below the expected value. Experimentally, an early work by Woollam \textit{et. al.} studied the insertion of K/Na atoms into bulk MoS$_2$, which found the maximum superconducting $T_{\rm C}$'s of K$_{x}$MoS$_2$ and Na$_x$MoS$_2$ to be about $7$ K and $3.2$ K respectively. \cite{woollam1976superconducting} In this sense, the electron doped MoS$_2$ should be a promising superconductor although its real $T_{\rm C}$ is seriously suppressed by real doping methods. Very recently, the dynamical stability and superconductivity have been reported in the free-standing Li-intercalated MoS$_2$.\cite{Huang2016} However, in real situation these 2D few layers are putted on particular substrates and will be affected by the lattice mismatch, which have not been addressed in Ref.~\onlinecite{Huang2016}.

In this work, the lattice dynamics and electron-phonon (EP) coupling of Na-intercalated MoS$_2$ and MoSe$_2$ bilayers have been studied via first-principles density functional theory (DFT) and density function perturbation theory (DFPT). Our calculations confirm the superconductivity in MoS$_2$ and MoSe$_2$ with Na intercalation. More interestingly, this superconductivity can be significantly enhanced by tensile strain. Our calculations will motivate more experimental studies to tune the physical properties like superconductivity of 2D materials by the strain effect.

\section{Model \& methods}
The DFT calculations have been performed using the PWSCF program of the Quantum-ESPRESSO distribution. \cite{giannozzi2009quantum} The ultrasoft pseudo-potential (including the semicore electrons as valence electrons in case of Mo) and generalized gradient approximation of Perdew-Burke-Ernzerhof (GGA-PBE) are used with a cutoff energy 35 Ry for the expansion of the electronic wave function in the plane waves. The vdw interactions are treated using the (Grimme) DFT-D2 approximation. \cite{grimme2006semiempirical}

Mo$X_2$ bilayers are modeled using slabs with one Na layer (one Na per one unit cell area) inserted. The surfaces are simulated by adding a vacuum layer of $\sim15$ {\AA}. For the electronic structure calculations, the Brillouin zone (BZ) integrations are performed with an $18\times18\times1$ grid by using the first-order Hermite-Gaussian smearing technique. Within the framework of the linear response theory, the dynamical matrices are calculated for $6\times6\times1$ grid of special $q$ points in the irreducible two-dimensional BZ and are Fourier interpolated throughout the full Brillouin zone. The dense $36\times36\times1$ grid is used in the BZ integrations in order to produce the accurate electron-phonon (EP) interaction matrices.

\section{Results \& discussion}
\subsection{Crystalline \& electronic structures}
Although the stablest Mo$X_2$ bilayers are stacked as the A-B type (Fig.~\ref{Fig1}(a-b)),\cite{liu2012tuning} this type of stacking conformation becomes dynamic unstable when Na monolayer is interacted into the Mo$X_2$ bilayers, as evidenced by the imaginary frequencies of phonon spectrum around $\Gamma$ point. Alternatively, the A-A type stacking conformation (Fig.~\ref{Fig1}(c-d)) is dynamic stable, which will be further discussed in Sec.III.B. Similar conclusion was also recently reported in Li-interacted MoS$_2$ bilayer.\cite{Huang2016} Therefore, considering the dynamic stabilization, the conformation as sketched in Fig.~\ref{Fig1}(d) will be systematically studied in the following. Each primitive cell contain two Mo$X_2$ layers and one Na atom.

First, the relaxation is performed until the force on each atom is smaller than $10^{-4}$ Ry/a.u.. The optimized lattice constant and interatomic distances are listed in Table~\ref{Table1} in comparison with the experimental values. The calculated lattice constant is only slight larger than the experimental value, which is quite reasonable since GGA normally overestimates lattice constants.

\begin{figure}
\centering
\includegraphics[width=0.48\textwidth]{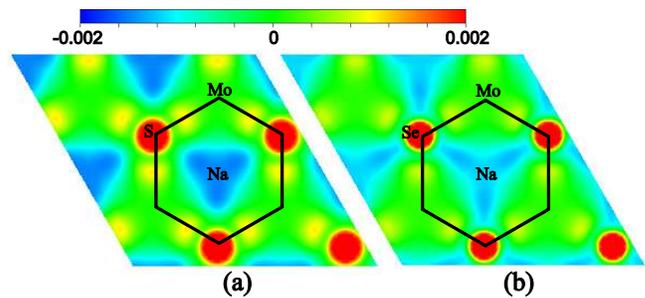}
\caption{(Color online) Electron density difference viewed along the ($001$) direction for (a) (MoS$_2$)$_2$Na and (b) (MoSe$_2$)$_2$Na.}
\label{Fig2}
\end{figure}

The calculated electron density difference are visualized in Fig.~\ref{Fig2}, which indicates the spatial distribution difference of electron density between the pure Mo$X_2$ bilayer and Na-intercalated Mo$X_2$ bilayer. The charge transfer from Na to $X1$ is obvious, especially for $X$=S, which changes the vdw force between original bilayers to ionic-bond-like interactions between Na and $X1$. As a direct result, the layer distance ($d$) is significantly shorten, as listed in Table~\ref{Table1}.

\begin{table}
\centering  % ±í¾ÓÖÐ
\caption{The optimized structural parameters (as defined in Fig.~\ref{Fig1}) in unit of {\AA}. The experimental values (with the superscript $E$) are also listed. \cite{boker2001band,kulikov1992intercalation}}
\begin{tabular*}{0.48\textwidth}{@{\extracolsep{\fill}}lcccccccc}
\hline
\hline
 & MoS$_{2}^E$ & MoS$_2$ & (MoS$_2$)$_2$Na & MoSe$_2^E$ & MoSe$_2$ & (MoSe$_2$)$_2$Na \\
\hline
$a$ & 3.161 & 3.204 & 3.251 & 3.285 & 3.330  & 3.385 \\
$h$ & 3.072 & 3.119 & 3.130 & 3.225 & 3.344  & 3.346 \\
$d$ & 3.074 & 3.113 & 2.249 & 3.225 & 3.178  & 2.435 \\
\hline
\hline
\end{tabular*}
\label{Table1}
\end{table}

\begin{figure}
\centering
\includegraphics[width=0.48\textwidth]{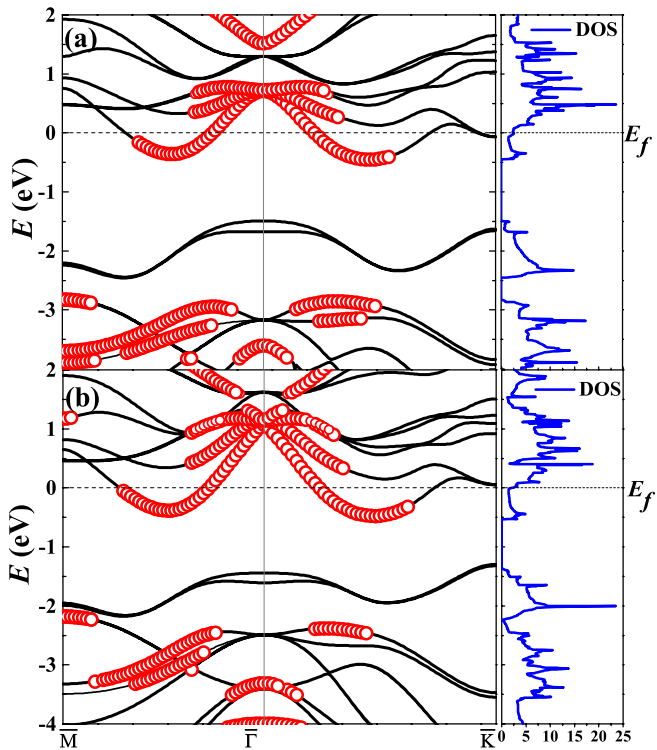}
\caption{(Color online) Electronic band structures and DOS's. The red circles indicate those states with more than $50\%$ of density contributed by Na and $X1$. (a) (MoS$_2$)$_2$Na; (b) (MoSe$_2$)$_2$Na$_2$.}
\label{Fig3}
\end{figure}

The calculated band structures and density of states (DOS) are shown in Fig.~\ref{Fig3}. As expected, the Fermi level crosses the conduction band as a direct result of Na intercalation, which dopes electrons to $X1$'s $p$ orbitals and builds strong ionic bonding between Na-$X$. It should be noted that this intercalation is different from previously studied Na-adsorption on MoS$_2$ by Komesu \textit{et.al.},\cite{komesu2014occupied} who found a semiconductor behavior with a narrowed band gap.

%The pristine bilayer MoS$_2$ and MoSe$_2$ are semiconductor with indirect gap about 1.29eV and 1.46eV respectively \cite{PRB385-033305}.
\subsection{Phonon, electron-phonon coupling, \& superconductivity}
\begin{figure}
\centering
\includegraphics[width=0.48\textwidth]{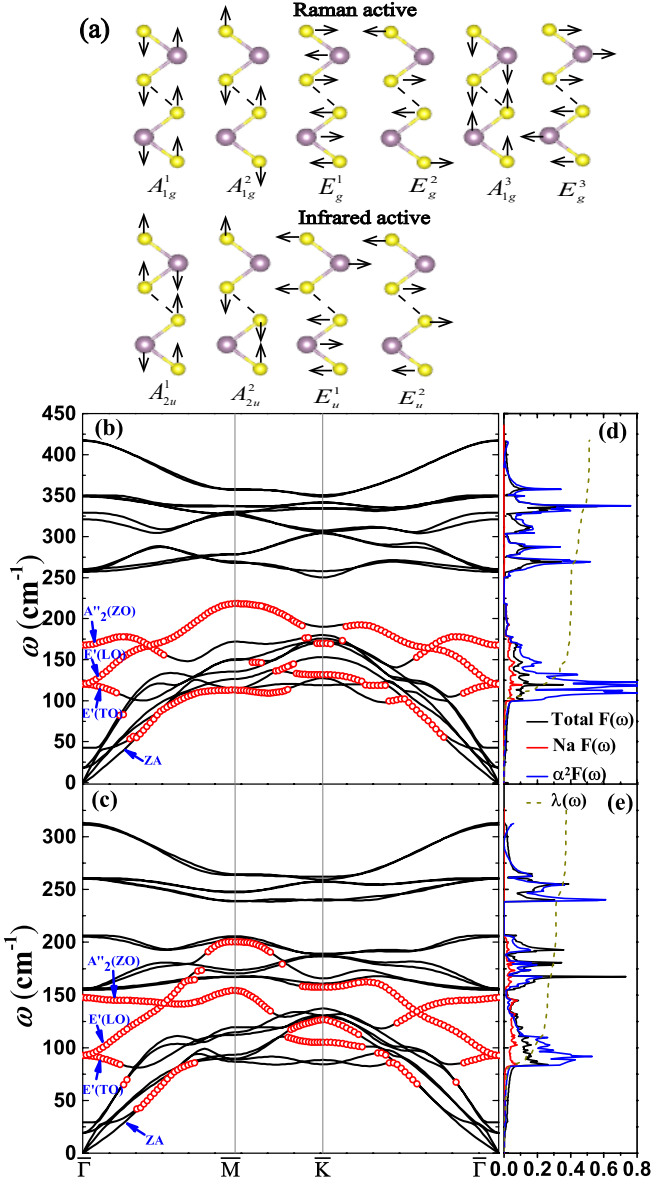}
\caption{(Color online) (a) Sketch of vibration modes of Mo$X_2$ bilayer. (b-c) Calculated phonon dispersion, phonon DOS, electron-phonon coupling $\lambda$, and Eliashberg spectral function for (MoS$_2$)$_2$Na system [(b) and (d)] and for (MoSe$_2$)$_2$Na system [(c) and(e)].}
\label{Fig5}
\end{figure}

%After electronic structure properties, we study the phonon behavior and electron-phonon coupling of (MoS$_2$)$_2$Na and (MoSe$_2$)$_2$Na from density functional perturbation theory.
Then we turn to pay attention to the phonon modes at the zone center. The pure MoS$_2$ and MoSe$_2$ bilayers have the identical point group ($D_{3d}$), which is reduced to $D_{3h}$ in (Mo$X_2$)$_2$Na. For the $D_{3d}$ point group, the optical modes at BZ center $\Gamma$ point can be decomposed as $3A_{1g}\oplus 2A_{2u}$ polarized along the hexagonal $c$ axis direction and $3E_{g}\oplus 2E_{u}$ polarized in the hexagonal closed packed plane. The phonon modes $E_{g}$ and $A_{1g}$ are both Raman (R) active, while the $A_{2u}$ and $E_{u}$ modes are infrared (IR) active, as sketched in Fig.~\ref{Fig5}(a). In contrast, for the $D_{3h}$ point group, the optical modes at $\Gamma$ point can be decomposed as $A''_{2} \oplus A'_{1} \oplus E' \oplus E''$. $A''_{2}$ and $E''$ are infrared active, while $A'_{1}$ and $E'$ are Raman active. The calculated frequencies of these optical modes are listed in Table~\ref{Table2}, in comparison with some experimental data.

\begin{table}
\centering
\caption{The calculated frequencies (in unit of cm$^{-1}$) of vibratory modes at $\Gamma$ point for (Mo$X_2$)$_2$Na system and pure Mo$X_2$ bilayer. The experimental values of $A_{1g}^2$ ($A_{2u}^2$) and $E_{g}^1$ ($E_{u}^1$) for pure MoS$_2$ (MoSe$_2$) bilayer are $406.1$ cm$^{-1}$ ($242.8$ cm$^{-1}$) and $384.9$ cm$^{-1}$ ($287.1$ cm$^{-1}$), respectively.\cite{chen2015helicity}}
\begin{tabular}{@{}lcccccc@{}}
\hline
\hline
R   &$A_{1g}^1$  &$A_{1g}^2$ &$E_{g}^1$  &$E_{g}^2$  &$A_{1g}^3$ &$E_{g}^3$\\
\hline
MoS$_2$ & 464.3 &398.9   &376.7  &280.9  &25.6 &20.3\\
MoSe$_2$& 345.1  &237.6 &277.6  &163.7 &36.2 &18.7\\
\hline
R    &$A_{1}'(1)$ &$A_{1}'(2)$ &$E'(1)$  &$E'(2)$  &$A_{1}'(3)$ &$E'(3)$\\
\hline
(MoS$_2$)$_2$Na   &416.9 &329.0 &349.3  &260.6  &39.0 &15.9\\
(MoSe$_2$)$_2$Na  &313.2 &206.3 &260.4   &154.8 &29.5 &19.2\\
\hline
IR   &$A_{2u}^1$ & $A_{2u}^2$  &$E_{u}^1$  &$E_{u}^2$ &$~$ &$~$\\
\hline
MoS$_2$ &462.6 &405.4 &376.7  &280.2 &$~$ &$~$\\
MoSe$_2$ &344.1 &235.7   &277.1  & 163.1 &$~$ &$~$\\
\hline
IR    &$A_{2}''(1)$ &$A_{2}''(2)$  &$E''(1)$  &$E''(2)$ &$~$ &$~$\\
\hline
(MoS$_2$)$_2$Na    &416.5 &321.3 &349.3  &257.1 &$~$ &$~$\\
(MoSe$_2$)$_2$Na   &311.7 &205.2 &260.6  &156.0 &$~$ &$~$\\
\hline
\hline
\end{tabular}
\label{Table2}
\end{table}

For pure Mo$X_2$ bilayers, the calculated phonon frequencies are only slightly smaller than the corresponding experimental values.\cite{chen2015helicity} This tiny inaccuracy is understandable since the GGA-PBE pseudo-potential normally overestimates the volume of cell, which softens the phonon modes. For (Mo$X_2$)$_2$Na systems, there are considerable red shifts of phonon frequencies compared to the corresponding ones of pure Mo$X_2$ bilayers except $A_{1}'(3)$ for (MoS$_2$)$_2$Na and $E'(3)$ for (MoSe$_2$)$_2$Na. On one hand, the expanded lattice structure by the intercalated Na layer (e.g. see $a$'s and $h$'s in Table~\ref{Table1}) leads to weaker force constants between $X$-Mo, which softens phonon modes. In fact, due to a larger lattice constant of (MoSe$_2$)$_2$Na compared with (MoS$_2$)$_2$Na, the frequencies of phonon modes in (MoSe$_2$)$_2$Na are correspondingly smaller than those of (MoS$_2$)$_2$Na. On the other hand, the aforementioned charge transfer from Na to $X1$ makes the $X1$-Na links be strongly ionic-type, which may suppress the neighbor covalent $X$-Mo bonds as a side effect. The $A'(3)$ and $E'(3)$ mode, which are layer and shear breathing modes respectively, are sensitive to the interlayer interaction. Due to the electronegativity difference between S and Se, more (less) charge is transferred from Na to neighboring S (Se), making stronger (weaker) Coulomb attraction between Na monolayer and MoS$_2$ (MoSe$_2$) layers. Thus, the changes of $A'(3)$ and $E'(3)$ modes in Na(MoSe$_2$)$_2$ are analogous to the trilaminar MoSe$_2$ case,\cite{chen2015helicity} i.e. $A'(3)$ is softening and $E'(3)$ is stiffening. In contrast, the strong Coulomb attraction between Na monolayer and MoS$_2$ layer moves these two modes toward opposite directions, i.e. $A'(3)$ is blue shifted and $E'(3)$ is red shifted, similar to the Li-intercalated MoS$_2$ case.\cite{Huang2016}

The calculated phonon dispersions along major high symmetry lines and phonon densities of states (PDOS, $F$($\omega$)) for (Mo$X_2$)$_2$Na are shown in Fig.~\ref{Fig5}(b-e). No imaginary frequency exists in the full phonon spectra, indicating the dynamical stability of the calculated structures of (Mo$X_2$)$_2$Na. %We can see that the extended range of phonon frequency decreases from (MoS$_2$)$_2$Na to (MoSe$_2$)$_2$Na. This is as expected from the change of lattice constants.
Since Na's vibration modes own the identical symmetry to these $X$-Mo-$X$ ones, the phonon eigenvectors have a strongly mixed character of Na atom and Mo$X_2$ triatomic layer, as indicated in Fig.~\ref{Fig5}(b-c). %In order to clearly the effect of the Na intercalation on phonon spectra of MoS$_2$ and MoSe$_2$, the Na atoms make a greater than 50\% contribution to phonon modes which similar to the band structure are plotted by open circles in Fig.~\ref{Fig5}(b) and Fig.~\ref{Fig5}(c).
It is clear that the vibrations contributed by Na are in the intermediate- and low-frequency region, as well as the out-of-plane acoustic mode (ZA). Generally, the acoustic modes and the layer-breathing modes for opposite vibrations of two triatomic layers are in the low-frequency range and the sandwich $X$-Mo-$X$ bond-stretching modes are in the high-frequency range due to the strong covalent bonding.\cite{huang2015prediction} %For (MoS$_2$)$_2$Na (Fig.~\ref{Fig5}(b)), the high-frequency region has few Na atom contribution, although the Na atom has a near equivalent mass to S atoms, which show the strong covalent bonding between S atoms and Mo atoms in MoS$_2$ triatomic layer. The triatomic layer breathing modes is softening since the charges transfer from Na atom to S1 which have been discussed above.
As shown in Fig.~\ref{Fig5}(b), two interlayer-shear modes [$E'$(TO): optical in-plane transverse mode, $E'$(LO): optical in-plane longitudinal mode] and one interlayer-breathing mode [$A_{2}''$(ZO): optical out-plane mode] are highly mixed with the Na layer's contribution, especially for the $E'$(LO) mode. And due to the moderate Coulomb attraction between Na and neighboring $X1$ ions, these $E'$(TO), $E'$(LO), and $A_{2}''$(ZO) modes situate in the intermediate frequency region. %For (MoSe$_2$)$_2$Na Fig.~\ref{Fig5}(c)), the conclusion is kinder different due to the Se atom mass is two times larger than S atom mass. Therefore, Na atom contribution extend to medium-high frequency (Fig.~\ref{Fig5}(c) and (e)).

In the following, the EP interaction is estimated. According to the Migdal-Eliashberg theory, the Eliashberg spectral function [$\alpha^{2}F(\omega)$] is given by:\cite{grimvall1981electron}
\begin{eqnarray}
\alpha^2F(\omega)&=&\frac{1}{N(E_{f})}\sum_{\nu}\sum_{kqjj'}\left | g_{(k+q)j'kj}^{q\nu} \right |^{2} \nonumber\\
 && \delta(\varepsilon_{kj}-\varepsilon_{F})\delta(\varepsilon _{(k+q)j}-\varepsilon_{F})\delta (\omega-\omega_{q\mu})
\end{eqnarray}
where $N(E_f)$ is the electronic DOS at Fermi level; $g_{(k+q)j'kj}^{q\nu}$ is the EP matrix element which can be determined self-consistently by the linear response theory. The EP coupling coefficient $\lambda$ is obtained by evaluating:\cite{PhysRev.187.525}
\begin{equation}
\lambda=2\int_{0}^{\infty}\frac{\alpha^{2}F(\omega)}{\omega}d\omega.
\label{lambda}
\end{equation}
The calculated coefficents are summarized in Table~\ref{Table3} and the Eliashberg functions for (Mo$X_2$)$_2$Na are shown in Fig.~\ref{Fig5}(c-d). The similarity between $F(\omega)$ and $\alpha^2F(\omega)$ indicates that all vibration modes contribute to the EP interaction. However, those high-frequency phonons do not contribute much to the strength of electron-phonon interaction due to the weighting of $1/\omega$ in the definition of $\lambda$ (see Eq.~\ref{lambda}).

According to Fig.~\ref{Fig5}(d-e), it's obvious that $E'$(TO), $E'$(LO), and $A_{2}''$(ZO) modes in (Mo$X_2$)$_2$Na make great contribution to $\lambda$  by increasing $\alpha^{2}F(\omega)$ curve peak in the low frequency region. As summarized in Table~\ref{Table3}, $\lambda$ is larger in (MoS$_2$)$_2$Na than in (MoSe$_2$)$_2$Na. The physical reasons are: 1) Larger DOS value at the Fermi level in (MoS$_2$)$_2$Na; 2) The contribution from $E'$(TO), $E'$(LO), and $A_{2}''$(ZO) modes are stronger in (MoS$_2$)$_2$Na.

\begin{table}
\centering
\caption{Calculated superconducting
 $T_{\rm C}$ in unit of K, electron-phonon coupling $\lambda$, logarithmically averaged frequency $\omega _{ln}$ (in unit of K) and electronic DOS at Fermi level $N(E_f)$ (states/eV).}
\label{Table3}
\begin{tabular*}{0.48\textwidth}{@{\extracolsep{\fill}}lcccccccc}
\hline
\hline
    & $N(E_f)$ & $\omega_{ln}$ & $\lambda$ & $T_{\rm C}$ \\
\hline
(MoS$_2$)$_2$Na  & $2.377$ & $219.452$ & $0.509$ & $2.858$ \\
(MoSe$_2$)$_2$Na & $2.143$ & $177.137$ & $0.373$ & $0.484$  \\
\hline
\hline
\end{tabular*}
\end{table}

The superconducting $T_{\rm C}$ can be estimated using the Allen-Dynes modified McMillan equation:\cite{allen1975transition}
\begin{equation}
T_{\rm C}=\frac{\omega_{ln}}{1.2}\exp[-\frac{1.04(1+\lambda)}{\lambda-\mu^*(1+0.62\lambda)}],
\end{equation}
where $\mu^*$ is the Coulomb repulsion parameter and $\omega_{ln}$ is the logarithmically averaged frequency. When taking a typical value $\mu^{*}=0.1$, the calculated $T_{\rm C}$ is $2.858$ K for (MoS$_2$)$_2$Na, which is very close to the measured values (about $2.2$-$3.2$ K) for Na-doped MoS$_2$ bulk.\cite{woollam1976superconducting} However, the obtained $T_{\rm C}$ ($0.484$ K) for (MoS$_2$)$_2$Na is very low. %Through the above analysis, we can see that Na atom vibration modes highly mix with $E'$(TO), $E'$(LO) and $A_{2}''$(ZO) modes which lead to improve $\lambda$ and higher DOS at the Fermi level (E$_{f}$) play an important role in $T_{\rm C}$.

The substrate strain from lattice mismatch is a widely used method to tune the physical properties of 2D materials, e.g. zero-field quantum Hall effect in graphene.\cite{guinea2010energy} Then it is interesting to study the lattice mismatch effect to the superconducting $T_{\rm C}$ of  Na-intercalated Mo$X_2$.

\begin{figure}
\centering
\includegraphics[width=0.48\textwidth]{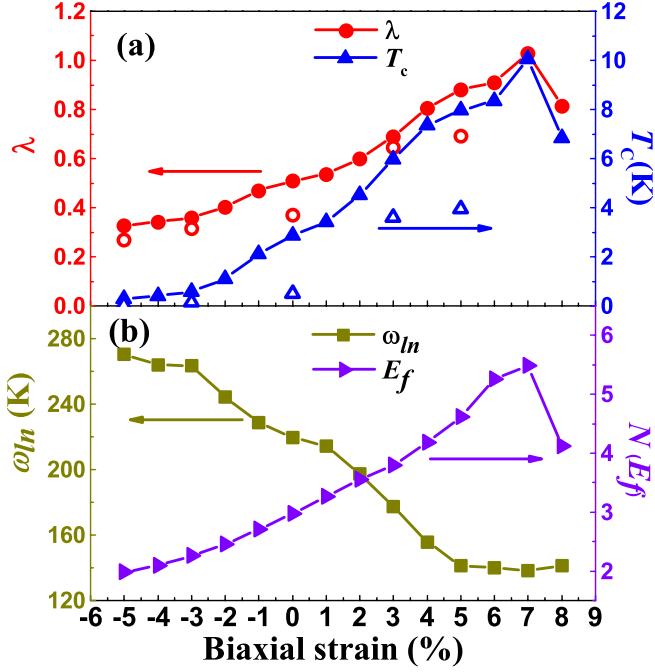}
\caption{(Color online) Calculated biaxial stress effects for (Mo$X_2$)$_2$Na (solid symbols: (MoS$_2$)$_2$Na; open symbols: (MoSe$_2$)$_2$Na). The dynamic stability for both cases persists within this strain region. (a) Superconducting $T_{\rm C}$ (right axis) and electron-phonon coupling $\lambda$ (left axis). (b) Logarithmically averaged frequency $\omega _{ln}$ and electronic DOS at Fermi level $N(E_{f})$. }
\label{Fig6}
\end{figure}

The biaxial strain ($\epsilon$) is imposed to simulate the lattice mismatch and the results are shown in Fig.~\ref{Fig6}. Interestingly, for (MoS$_2$)$_2$Na with increasing tensile strain, $T_{\rm C}$ continuously increases to a maximum value $10.049$ K at $\epsilon=+7\%$, beyond which $T_{\rm C}$ turns to decrease. In contrast, the compressive strain can suppress $T_{\rm C}$ monotonously, e.g. to $0.295$ K when biaxial strain down to $\epsilon=-5\%$.

Physically, a tensile strain suppresses the bilayer thickness and thus shortens the distance between Na and S1, which can result in more electron transfer from Na to S1 and improve the DOS at the Fermi level (as shown in Fig.~\ref{Fig6}(b)). In addition, tensile strain also leads to stronger electron-phonon interaction by softening the $E'$(TO), $E'$(LO), and $A_{2}''$(ZO) modes. Similar conclusion is also reached in (MoSe$_2$)$_2$Na whose $T_{\rm C}$ is enhanced up to $3.944$ K for $\epsilon=+5\%$ biaxial tensile strain but suppressed to near zero for compressive strain.

Previous studies suggested that in general the superconducting $T_{\rm C}$ of 2D materials could be improved by increasing doping density.\cite{xue2012superconductivity,ge2013phonon,huang2015prediction} Our calculation gives one more route to improve $T_{\rm C}$ effectively and easily.

\subsection{Phonon mean free path}
Finally, in order to deeply understand the lattice dynamics of Na-intercalated Mo$X_2$, the phonon mean free path (MFP) is also investigated, which now can be directly measured in experiments.\cite{minnich2011thermal} The phonon MFP for the mode at $q$ point with $s$ polarization is defined as $Q_{qs}=V_{qs}\tau_{qs}$, where $V_{qs}$ is the phonon group velocity and $\tau_{qs}$ is phonon relaxation time. In real materials, the phonon relaxation time is determined by the combination of electron-phonon, phonon-phonon umklapp scattering, and boundary scattering. For few-layer 2D materials, the boundary scattering from the sides should be very weak and negligible.\cite{wang2011absence} For simplify, a uniform lifetime ($\tau^{ep}$) is assumed for all phonon modes' contributions to the electron-phonon scattering, and the Debye spectrum is adopted as the phonon density of state approximatively \cite{souvatzis2011phonon}. Then, $\alpha^2F(\omega)$ can be approximated as:\cite{souvatzis2011phonon}
\begin{equation}
\alpha^2F(\omega)\approx\frac{3(\tau^{ep})^{-1}}{2\pi N(E_f)}\frac{\omega}{\hbar\omega_{\rm D}^3},
\end{equation}
where $\omega_{\rm D}$ is the Debye frequency. Corresponding the $\lambda$ can be integrated as:
\begin{equation}
\lambda=2\int_0^\infty\frac{\alpha^{2}F(\omega)}{\omega}d\omega=\frac{6(\tau^{ep})^{-1}}{2\pi N(E_f)}\frac{1}{\hbar\omega_{\rm D}^2}.
\end{equation}
Then using the value of $\lambda$ calculated before, the effective $\tau^{ep}$ can be estimated, as presented in Table~\ref{Table4}.

The three-phonon umklapp scattering lifetimes ($\tau^{ph-ph}$) for different phonon branches can be estimated as:\cite{klemens2000theory,klemens2001theory}
\begin{equation}
\tau_{qs}^{ph-ph}=(2\gamma_{qs}^2\frac{k_BT}{Mv_s^2}\frac{\omega_{qs}^2}{\omega_{\rm D}})^{-1},
\end{equation}
where $M$ is the mass of a (Mo$X_2$)$_2$Na unit cell, $v_s$ is the average phonon velocity for a given branch, $T$ is temperature, $k_B$ is the Boltzmann constant, and $\gamma$ is the Gr\"{u}neissen parameter. For 2D materials, the $\gamma$ of each phonon mode at $q$ point with $s$ polarization is given by:\cite{zou2001phonon,mounet2005first}
\begin{equation}
\gamma_{qs}=-\frac{a}{2\omega_s(q)}\frac{d\omega_s(q)}{da}
\end{equation}

\begin{table}
\centering
\caption{The calculated Debye frequency $\omega_{\rm D}$, phonon-phonon relaxation time ($\tau^{ph-ph}$), electron-phonon relaxation time ($\tau^{ep}$), and MPF for (Mo$X_2$)$_2$Na system around the $\Gamma$ point at room temperature and at the frequency of 4 cm$^{-1}$.}
\label{Table4}
\begin{tabular*}{0.48\textwidth}{@{\extracolsep{\fill}}lcccccccc}
\hline
\hline
   & $\omega_{\rm D}$ (cm$^{-1}$) & $\tau^{ph-ph}$ (ps) & $\tau^{ep}$ (ps) & MPF (nm) \\
\hline
(MoS$_2$)$_2$Na  & $237.0$     &$237.4$ (TA)    &$23.8$    &$142.4$ (TA)\\
 $~$      & $~$    &236.8 (LA)   &$~$       &$229.9$ (LA)\\
\hline
(MoSe$_2$)$_2$Na    &$171.9$     &$1077.2$ (TA)   &$68.5$    &$536.2$ (TA)\\
$~$   &$~$       &$913.6$ (LA)    &$~$       &$612.3$ (LA)\\
\hline
\hline
\end{tabular*}
\end{table}

Since the acoustic modes (particularly the LA and TA modes) have relatively larger velocities around the $\Gamma$ point than those of the optical modes, they contribute to most of $Q$.\cite{ong2011effect} Here $\gamma_{qs}$ is averaged around $\Gamma$ point. The calculated MFP's of acoustic modes around the $\Gamma$ point are listed in Table~\ref{Table4}. The phonon-phonon umklapp scattering makes about $90\%$ contributions to MFP for (Mo$X_2$)$_2$Na. The anharmonic terms in lattice vibration have great influences on MFP and mainly reflect in $\gamma$ here. Generally, larger $\gamma$ leads to smaller MFP. For (MoS$_2$)$_2$Na, we obtain $\gamma$ for TA and LA modes are $2.64$ and $2.11$ respectively, while in (MoSe$_2$)$_2$Na are only $1.47$ and $1.34$ respectively. The stronger anharmonicity in (MoS$_2$)$_2$Na leads to smaller MFP. Therefore, the MFP's ($536.2$ nm for the TA branch, $612.3$ nm for the LA branch) in (MoSe$_2$)$_2$Na is quite prominent, approaching that of graphene (about $775$ nm).\cite{ghosh2008extremely} Even for (MoS$_2$)$_2$Na, the MFP's ($142.4$ nm for the TA branch and $229.9$ nm for the LA branch) are larger than those of monolayer MoS$_2$ (about $103.1$ nm for the LA branch calculated using the same method). While in Ref.~\onlinecite{cai2014lattice}, the MFP for the LA branch was reported to be only $18.1$ nm for MoS$_2$ monolayer, which was partially due to the inaccuracy $\gamma$ in their calculation, as pointed out by Ref.~\onlinecite{huang2014correlation,sevik2014assessment}. In short, the Na-intercalation can significantly improve the  phonon mean free path of (Mo$X_2$)$_2$Na, which may be used in heatconduction devices.

\section{Conclusion}
An in-depth understanding of electronic properties, the lattice dynamical properties, and superconductivity of modificatory 2D materials is highly important for its potential applications in heatconduction devices as well as nanoscale superconductor. The present DFT study found that the Na atoms intercalation can effectively change the electronic properties and lattice dynamical properties of MoS$_2$ and MoSe$_2$.

In summary, electrons transfer from intercalated Na atoms to neighboring Se or S atoms, which increase the density of states at the Fermi level and a semiconducting-to-metallic transition. The superconductivity is expected to be induced by such Na-intercalation, and the superconducting $T_{\rm C}$ would be enhanced by biaxial tensile strain. In addition, the phonon mean free path at room-temperature is also greatly improved in Na intercalated MoSe$_2$, which is advantaged for related applications.

\begin{acknowledgments}
Work was supported by National Natural Science Foundation of China (Grant Nos. 11274060 and 51322206).
\end{acknowledgments}

\bibliographystyle{apsrev4-1}
\bibliography{ref}
\end{document}